\documentclass[manuscript]{acmart}
\setcopyright{none}
\AtBeginDocument{%
  }

\usepackage{graphicx}
\usepackage{xurl}
\usepackage{xspace}
\def\aia{\textsc{\texttt{EU-AIA}}\xspace}

\usepackage{scalerel,graphicx,xparse}

\NewDocumentCommand\use{}{\scalerel*{\includegraphics{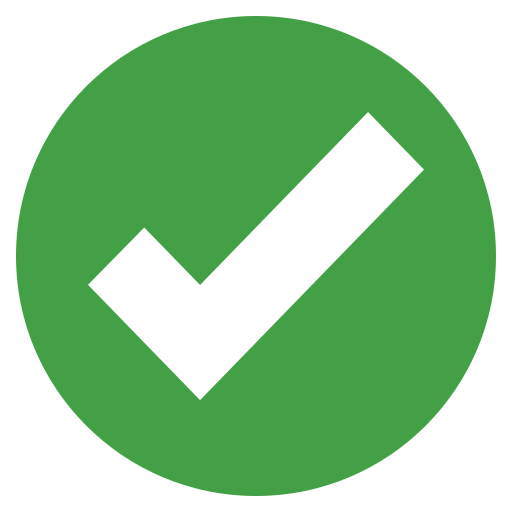}}{X}}
\NewDocumentCommand\risk{}{\scalerel*{\includegraphics{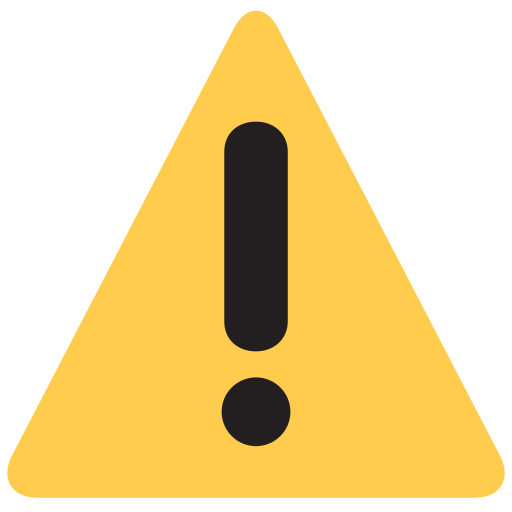}}{X}}
\NewDocumentCommand\rules{}{\scalerel*{\includegraphics{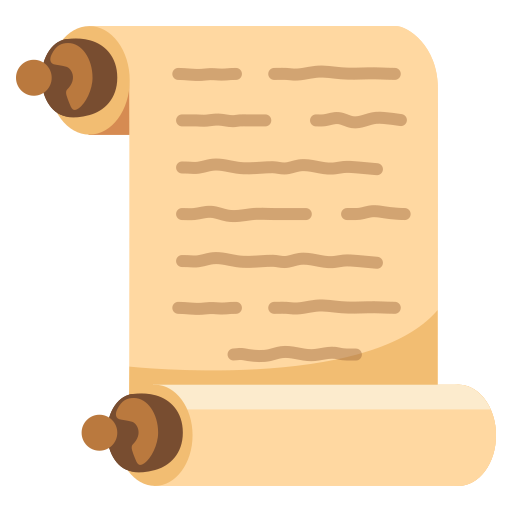}}{X}}
\NewDocumentCommand\lock{}{\scalerel*{\includegraphics{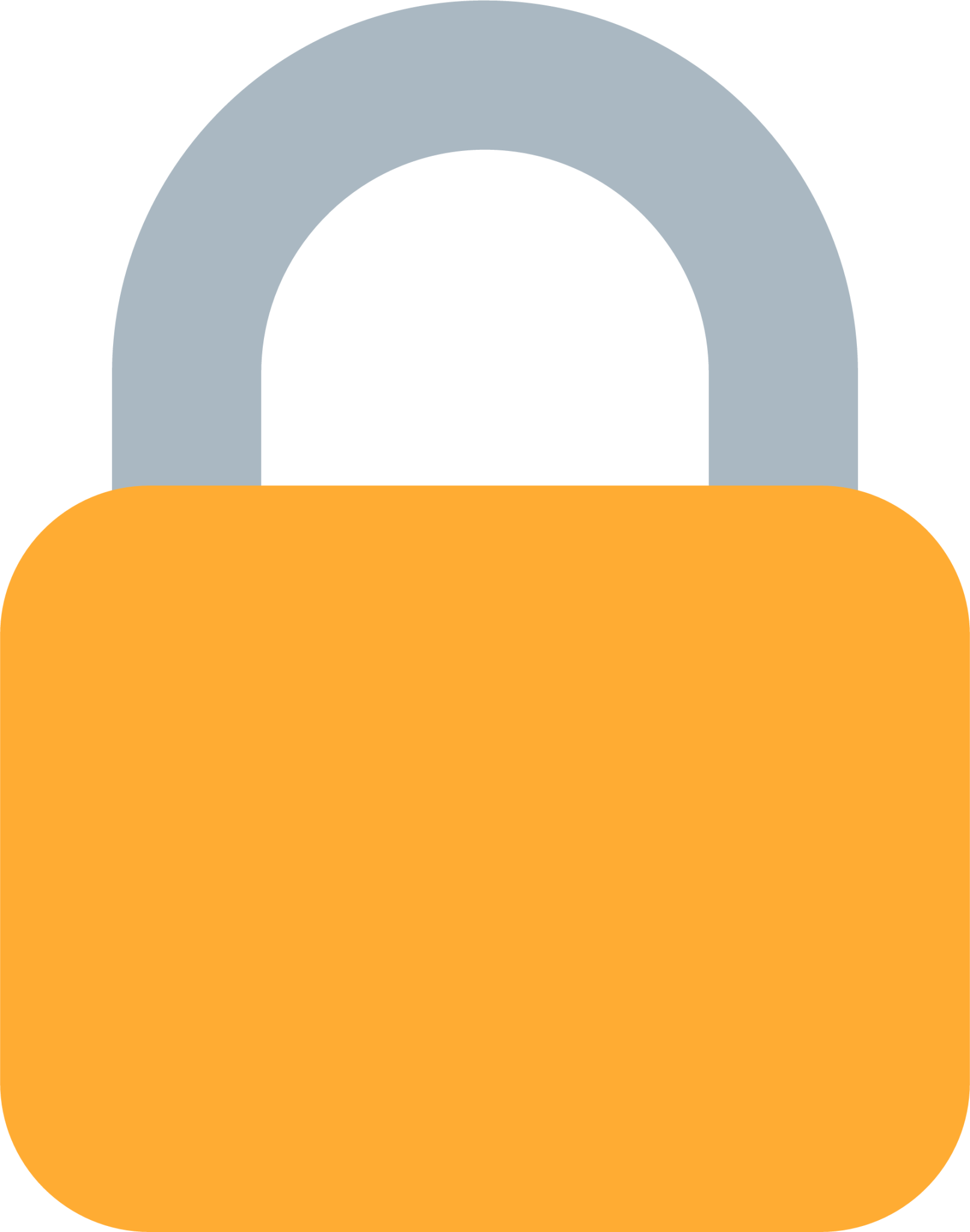}}{X}}
\NewDocumentCommand\student{}{\scalerel*{\includegraphics{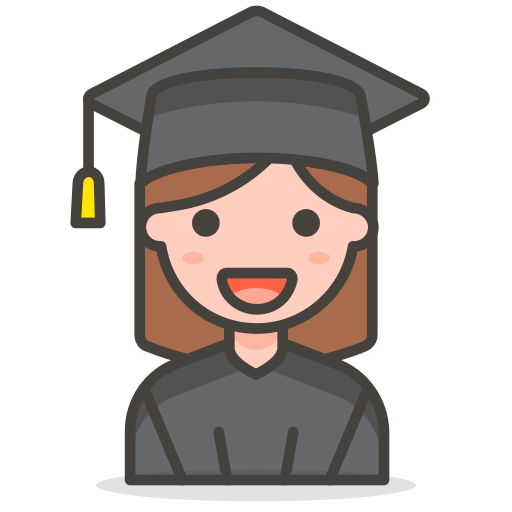}}{X}}
\NewDocumentCommand\newspaper{}{\scalerel*{\includegraphics{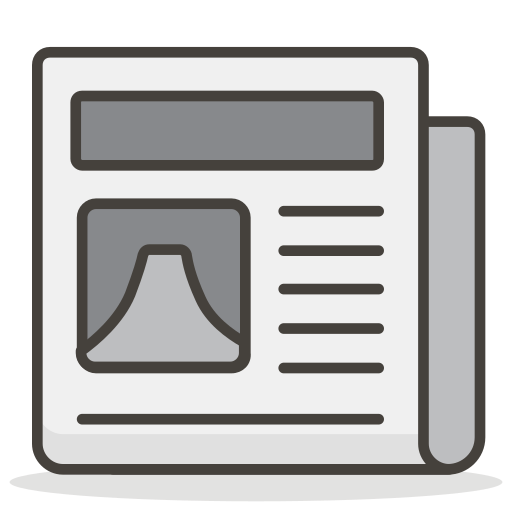}}{X}}

\setcopyright{acmlicensed}
\copyrightyear{2025}
\acmYear{2025}
\acmDOI{XXXXXXX.XXXXXXX}

\begin{document}

\title{AI Policy in Practice: Comparing Active Policies with International Legislation}
\title{AI Policy in Practice: Comparing Organisation Policies with Coming International Legislation (EU AI Act)}
\title{Local Differences, Global Lessons: Insights from Organisation Policies for International Legislation}

\author{Lucie-Aim\'{e}e Kaffee}
\authornote{Both authors contributed equally to this research.}
\affiliation{%
  \institution{Hugging Face}
  \country{\_}
  }
\email{lucie.kaffee@huggingface.co}

\author{Pepa Atanasova}
\authornotemark[1]
\affiliation{%
  \institution{University of Copenhagen}
  \country{Denmark}
}
\email{pepa@di.ku.dk}

\author{Anna Rogers}
\affiliation{%
  \institution{IT University of Copenhagen}
  \country{Denmark}
}
\email{arog@itu.dk}



\begin{abstract}
The rapid adoption of AI across diverse domains has led to the development of organisational guidelines that vary significantly, even within the same sector. This paper examines AI policies in two domains, news organisations and universities, to understand how bottom-up governance approaches shape AI usage and oversight. By analysing these policies, we identify key areas of convergence and divergence in how organisations address risks such as bias, privacy, misinformation, and accountability. We then explore the implications of these findings for international AI legislation, particularly the EU AI Act, highlighting gaps where practical policy insights could inform regulatory refinements. Our analysis reveals that organisational policies often address issues such as AI literacy, disclosure practices, and environmental impact, areas that are underdeveloped in existing international frameworks. We argue that lessons from domain-specific AI policies can contribute to more adaptive and effective AI governance at the global level. This study provides actionable recommendations for policymakers seeking to bridge the gap between local AI practices and international regulations.\looseness=-1
%
\end{abstract}




\received{20 February 2007}
\received[revised]{12 March 2009}
\received[accepted]{5 June 2009}

\maketitle

\section{Introduction}
The recent advancements in the performance of AI models\footnote{The policies discussed in this article mostly focus on the current systems commonly referred to as ``generative AI'', in that they can be used to generate synthetic texts/images etc. They are typically based on models of Transformer architecture pre-trained on large volumes of data (texts, images etc.). The exact definitions are rarely provided and are an active area of discussion in research (see references in \autoref{fig:sociotech-challenges}).} on a multitude of tasks, especially in zero-shot or few-shot scenarios \cite{KOCON2023101861, zhao2023survey}, have accelerated their adoption across different domains.  This rapid uptake has presented organisations with unprecedented challenges, necessitating the \textit{swift development of organisational guidelines} for the use of AI to manage associated risks and opportunities. These guidelines reflect bottom-up governance approaches tailored to local needs and operational contexts.

From the top-down perspective, the most significant governance effort currently is the EU AI Act (\aia) \cite{2024-ai-act} (\S\ref{sec:aia}), which provides overarching frameworks for managing high-risk AI systems. 
However, while such frameworks establish broad standards, their top-down approach often lacks the specificity required for effective implementation in diverse organisational settings. This creates gaps where organisations must independently interpret and address risks, resulting in guidelines that emphasise practical challenges, such as AI literacy, bias mitigation, and environmental sustainability, which are underdeveloped in existing legislation.
This raises our main research question: \textit{can the analysis of bottom-up initiatives (organisational guidelines and academic research) provide valuable insights into the areas that are not sufficiently addressed by the top-down framework of the \aia?}\looseness=-1

To address this question, this paper compiles a list of challenges known from AI research (\S\ref{sec:challenges}) and conducts case studies of organisational AI guidelines from two domains: universities and news organisations. Through iterative coding of guidelines developed by organisations in each domain (\S\ref{sec:method}), we examine discrepancies in how risks are classified, uses are prescribed, and the performance of AI models is perceived. We discuss the commonalities and discrepancies within each domain (news \S\ref{sec:policies:news}, universities \S\ref{sec:policies:uni}). 
We identify multiple areas in need of clarifications and further research, and provide 
%
actionable recommendations for policymakers 
on the topics of AI literacy training, digital inequity, disclosure of AI-generated content, bias, attribution, and environmental impact in the context of AI legislation (\S\ref{sec:insights}). Our findings underscore the potential of organisational policies and academic research to inform and refine global frameworks like the \aia. \looseness=-1

\section{Background}
\subsection{Sociotechnical Challenges for Organisations Known from AI Research}
\label{sec:challenges}

\begin{table}[t]
\begin{tabular}{p{11.7cm}p{2.7cm}}
\toprule
\textbf{Challenge} \& \textbf{Summary} & \textbf{Risk for the org.} \\
\midrule 
\textbf{What is regulated:} what kind of models even fall under the policy? Definitions can be based on training compute \cite{2024-ai-act}, data \cite{RogersLuccioni_2024_Position_Key_Claims_in_LLM_Research_Have_Long_Tail_of_Footnotes}, performance \cite{anderljungFrontierAIRegulation2023} etc. & Guidelines not scoped appropriately \\
\textbf{Detectability \& enforceability}: can we detect when AI models' usage violates the policy? Particularly, when generated content is used without disclosure? At present, no \cite{PuccettiRogersEtAl_2024_AI_News_Content_Farms_Are_Easy_to_Make_and_Hard_to_Detect_Case_Study_in_Italian}. & Guidelines not enforceable \\
\textbf{Factual errors}: the current models cannot reliably reject queries for which they do not have enough information \cite{amayuelas-etal-2024-knowledge}, and may output plausible-sounding but false results that are hard to identify and check \cite{zhang2023sirenssongaiocean,hicksChatgptBullshit2024}. Retrieval-augmented generation still has this problem \cite{mehrotraPerplexityBullshitMachine}. & Losing credibility and reputation \\
\textbf{Unsafe models}: in spite of attempts to force the models to follow certain content policies \cite{ouyang2022training}, the models can still violate them \cite{DerczynskiGalinkinEtAl_2024_garak_Framework_for_Security_Probing_Large_Language_Models}, and this training can even decrease the quality in some aspects \cite{10.1093/polsoc/puae020,casper2023open}. & Exposing employees to toxic outputs \\
\textbf{Privacy and security risks}: employees using non-local generative AI models may expose sensitive data from themselves and their organizations to the entity controlling such models \cite{Kim_2023_Amazon_warns_employees_not_to_share_confidential_information_with_ChatGPT_after_seeing_cases_where_its_answer_closely_matches_existing_material_from_inside_company} or third-party attackers \cite{WuZhangEtAl_2024_New_Era_in_LLM_Security_Exploring_Security_Concerns_in_Real-World_LLM-based_Systems}. Platform plugins may also increase vulnerabilities \cite{iqbal2024llm}. & Exposing sensitive data \\
\textbf{Misleading marketing claims}:  
employees may believe the claims of AI  model ``capabilities'' and trust the machine too much \cite{KheraSimonEtAl_2023_Automation_Bias_and_Assistive_AI_Risk_of_Harm_From_AI-Driven_Clinical_Decision_Support}, even though the benchmark results may be compromised by methodological problems and test set contamination \cite{RogersLuccioni_2024_Position_Key_Claims_in_LLM_Research_Have_Long_Tail_of_Footnotes}. 
& Degradation in the outputs of the organization \\
\textbf{Transparency \& accountability}: the social and legal norms on disclosing AI ``assistance'' and taking responsibility for the resulting text have not yet settled. The popular providers of these models do not accept responsibility for any faults in the output\footnote{}. & Public blame for any missteps \\
\textbf{Bias \& inequity}: The social biases in AI systems are well-documented  \cite{bolukbasi2016man,nadeem-etal-2021-stereoset,marchiori-manerba-etal-2024-social,stanczak2023quantifying,hutchinson-etal-2020-social,bender2021dangers,sharma2024generative}, 
and the use of such models may reinforce misrepresentations in the society. & Discrimination, ethical code violation \\
\textbf{Explainability}: checking model outputs would be easier if they were accompanied by rationales, the current interpretability methods are not faithful to the model’s actual decision-making \cite{lanham2023measuring,atanasova-etal-2023-faithfulness}. & Trusting unreliable solutions \\
\textbf{Brittleness}: Generative models perform worse outside of their training distribution \cite{McCoyYaoEtAl_2024_Embers_of_autoregression_show_how_large_language_models_are_shaped_by_problem_they_are_trained_to_solve,McCoyYaoEtAl_2024_When_language_model_is_optimized_for_reasoning_does_it_still_show_embers_of_autoregression_analysis_of_OpenAI_o1}. For language models, this includes changes in both language (idiolects, dialects, diachronic changes), content (e.g. evolving world knowledge), and slight variations in prompt formulation and examples \cite{zhu2023promptbench,LuBartoloEtAl_2022_Fantastically_Ordered_Prompts_and_Where_to_Find_Them_Overcoming_Few-Shot_Prompt_Order_Sensitivity}. & Employees wasting time and/or getting poor results \\
\textbf{Risks to creativity}: AI systems may generate unseen sequences of words, but their ``creativity'' is combinatorial, often lacking diversity, feasibility, and depth \cite{si2024can,padmakumar2024does}, and further degraded in languages other than English \cite{marco-etal-2024-pron}. Exposure to AI assistance could \textit{decrease} human creativity and diversity of ideas in non-assisted tasks \cite{kumar2024humancreativityagellms}. & Degradation in the outputs of the organization and its existing human resources \\
\textbf{Credit \& Attribution}: AI systems are commonly trained on copyrighted texts\cite{Gray_2024_OpenAI_Claims_it_is_Impossible_to_Train_AI_Without_Using_Copyrighted_Content} without author consent
. This practice triggered multiple lawsuits  \cite{BrittainBrittain_2023_Lawsuits_accuse_AI_content_creators_of_misusing_copyrighted_work,Vynck_2023_Game_of_Thrones_author_and_others_accuse_ChatGPT_maker_of_theft_in_lawsuit,JosephSaveriLawFirmButterick_2022_GitHub_Copilot_investigation,GrynbaumMac_2023_Times_Sues_OpenAI_and_Microsoft_Over_AI_Use_of_Copyrighted_Work,Panwar_2025_Generative_AI_and_Copyright_Issues_Globally_ANI_Media_OpenAI_TechPolicyPress}, protests from the creators \cite{Heikkila_2022_This_artist_is_dominating_AI-generated_art_And_hes_not_happy_about_it,More_than_15000_Authors_Sign_Authors_Guild_Letter_Calling_on_AI_Industry_Leaders_to_Protect_Writers}, and questions about the credit for the author of an ``assisted'' text \cite{FormosaBankinsEtAl_2024_Can_ChatGPT_be_author_Generative_AI_creative_writing_assistance_and_perceptions_of_authorship_creatorship_responsibility_and_disclosure}. & Legal exposure, violating plagiarism policies/principles \\
\textbf{Carbon emissions:} The current AI systems are environmentally costly for both training and inference \cite{luccioniCountingCarbonSurvey2023,dodge2022measuring,bouza2023estimate,liMakingAILess2023}, and workflows that significantly rely on them would have adverse effect on climate action. & Not meeting sustainability goals\\
\bottomrule
\end{tabular}
\caption{Major sociotechnical challenges for organizations relying on the current AI technology}
\label{fig:sociotech-challenges}
\end{table}
%
AI research literature point to numerous sociotechnical challenges for organisations relying on the current AI technology. Our work does not aim to provide a comprehensive survey, but we list the issues that we identified through literature review in \autoref{fig:sociotech-challenges}, together with the possible risks to organisations whose employees rely on this technology. This list serves as background to the types of problems that organisational policy or regulatory frameworks may identify as issues that need addressing.

\subsection{EU AI Act}
\label{sec:aia}
Likewise, the scope of this work does not allow for a detailed discussion of \aia, but for our purposes a key factor is that it implements a risk-based approach to regulating AI, in which systems are categorised systems by their potential threats. AI systems deemed ``unacceptable'' for their \textit{potential risks to EU values and fundamental rights}, such as AI systems performing social scoring, are outright banned. Systems considered ``high-risk'', including those used in critical infrastructure, law enforcement, employment, and education, face stringent ex-ante rules and are also subject to post-market monitoring. Universities, one of our case policy case studies, are in the education sector. Perhaps surprisingly, news is not identified as a high-risk application, given its possible consequences in election cycles. For general- and minimal-risk AI, the \aia primarily relies on transparency obligations, requiring that users be informed when they are interacting with or viewing outputs from certain AI systems.

There is also an important distinction between ``providers'' and ``deployers'', each bearing distinct responsibilities. Providers are entities that develop an AI model, place it on the market under their name or trademark (e.g., `Llama 3' by Meta), or substantially modify an existing AI model. They must ensure the model’s compliance with relevant standards, document design choices and data governance procedures, and, where required, undergo conformity assessments. If the model is a general purpose AI model (as Llama), it falls under additional regulations. Deployers, on the other hand, are the organisations or individuals that integrate and use the AI models in their systems (e.g. a university that creates an AI system using Llama to provide access to it to its staff and students). Their obligations typically focus on correct implementation, ensuring that the system is used within the scope of its intended purposes and monitoring its real-world performance for safety, accuracy, and potential harm.
In the case of universities, they can be both deployers of AI systems, if they offer their own AI systems, as well as end-users, subscribing to other AI deployers services, such as `ChatGPT'. 
For news organisations, in a majority of cases, the policies assume that they subscribe to an external AI deployer's systems.
Downstream users and large-scale distributors of AI generated content, as could be the case for news organisations, do not currently have obligations under the \aia.
In this article, we focus on the \aia and exclude discussions about related EU regulation, such as the \textit{Directive on Copyright in the Digital Single Market}.

\section{Related Work}

In recent years, both news organisations and educational institutions have proactively developed policies to guide the ethical and effective use of artificial intelligence (AI) within their respective domains. 

\textit{\textbf{News Organisations.}}
With the wide uptake of AI, news organisations have been impacted by the use and availability of AI models and systems, requiring them to create clear guidelines and policies on how to use this new technology \cite{simon2023ai}.
A comprehensive analysis of 52 news organisations across various countries reveals a concerted effort to address AI's implications on journalistic integrity, transparency, and accountability \cite{simon2023policies}.
Previous work has compared the transparency provisions in the European Union's AI Act, particularly Article 50, and their alignment with news readers' expectations \cite{piasecki2024ai}. Similar to our findings, the study highlights the necessity for clear disclosure when AI systems contribute to news content, as transparency is crucial for maintaining reader trust. 
\citet{de2022artificial} explore the current perceptions and future outlook of AI in news media, identifying key areas where AI technologies are being adopted, such as machine learning, computer vision, and natural language processing. While they find potential benefits of AI in enhancing news production and distribution, they also caution against challenges related to editorial standards and public trust.
In the context of visual AI in news organisations, \citet{thomson2024generative} emphasise the importance of clear guidelines to ensure clear boundaries between human-created and AI-generated images.
%

\textit{\textbf{Education.}}
The uptake of AI technologies also impacts educational Institutions, such as universities.
Studies point out a set of challenges with the use of ChatGPT in the educational institutions such as the misuse of ChatGPT to trick online exams \cite{susnjak2024chatgpt}; the higher correct answer rate on exams with the improvement of AI technologies \cite{de2023can}; the integration of generative AI technologies into engineering education given the limitations of the training data quality \cite{qadir2023engineering}; superficial or inaccurate responses, potentially misleading students and the risk of bias and discrimination \cite{farrokhnia2024swot}.

Hence, educational institutions are also actively formulating AI policies to navigate the integration of AI in academic settings. The 2024 EDUCAUSE Action Plan\footnote{\url{https://www.educause.edu/research/2024/2024-educause-action-plan-ai-policies-and-guidelines}} and UNESCO \cite{holmes2023guidance} outline comprehensive guidelines for AI adoption in higher education, both focusing on ethical considerations and the impact on teaching and learning practices.
\citet{ghimire2024guidelines} conducted a survey of academic institutions, such as high schools, to collect information on current policies w.r.t. generative AI. This study gives an insight into the current lack of AI policies in many education institutions in the US. 
\citet{slimi2023navigating} emphasise the importance of stakeholders working together to develop AI policies in the education space.
A set of studies have focused on the perspective of educators \cite{pischetola2024desirable}, proposing policy implications based on a survey of teachers \cite{chiu2023impact}.
%
%
\citet{dotan2024responsible} propose a ``points to consider'' approach for the responsible adoption of generative AI in higher education, emphasising alignment of AI integration with the unique goals, values, and structural features of higher education institutions.
%

\textit{\textbf{EU AI Act.}} \aia is the first attempt at a comprehensive international regulation of AI, but 
many important details are yet to be clarified in the \aia. Given the dynamic nature of AI development, the legislation will need updating over time \cite{cantero2024artificial}.
\citet{cantero2024artificial} highlight the critical role of standards in the \aia's co-regulation strategy, emphasising the need for reform to keep pace with the rapid evolution of AI technologies. 
Scholars have raised multiple concerns about the Act's approach, e.g., arguing that the EU conflates trustworthiness with the acceptability of risk \cite{laux2024trustworthy} and identifying limitations and loopholes in the \aia and related liability directives \cite{wachter2024limitations}. 
\citet{wachter2024limitations} criticise the \aia’s reliance on disclosure rather than substantive requirements, expressing concern that transparency and reporting alone will not adequately tackle systemic risks and societal harms. The legislation’s provisions on environmental impact also stop short of requiring actual reductions in energy consumption. Although the Act takes promising steps toward accountability, it falls short on deeper questions of fairness, acceptable risk, and enforceable standards. 
%
%

These critiques underscore the importance of analysing gaps in the \aia and pointing out possible improvements for future iterations, which is the purpose of this work. While previous studies have described organisational policies or analysed the gaps in the \aia, to the best of our knowledge, this is the first attempt to inform analysis of the gaps in the \aia by the existing and practically tested organisational policies as well as issues identified in academic research.\looseness=-1

\section{Methodology}
\label{sec:method}

\begin{table}[t]
\footnotesize

\begin{tabular}{p{2.6cm}p{11.5cm}}
\toprule
\textbf{Code} & \textbf{Description}\\
\midrule 
\multicolumn{2}{c}{\textbf{Suggested Uses}} \\
Illustrations/Graphics & \newspaper Using AI to generate illustrations and graphics for publication.\\
Image generation &  \newspaper Using AI to generate photorealistic images, \student visuals, diagrams, or other graphical representations.\\
Article generation &  \newspaper Using AI to generate long texts to be published as (part of) articles.\\
Data analysis & \newspaper Using AI to process, sort, and analyse large amounts of data.\\
Language tool &  \newspaper \student Using AI as a language assistant, e.g., for grammar and spelling checks, language proofreading, and rephrasing.\looseness=-1 \\
Transcription/Translation & \newspaper Transcription of, e.g., interviews and \student translating content between languages. \\ 
Ideas (Content) &  \newspaper Using AI for headlines, backgrounds, story ideas or sources to be approached.\\
Ideas (Marketing) & \newspaper Using AI for generating ideas for marketing campaigns.\\
Ideas (University) & \student Using AI to brainstorm and generate ideas, e.g., for project planning and curriculum development.\\ 
Content moderation &  \newspaper Using AI to filter spam, hate speech and fake news on social media or website comments.\\
Self-learning      & \student Supporting students' independent learning by providing alternative perspectives or personalised study plans\\ 
Course design      &  \student Generating learning objectives, developing rubrics, creating assessments, drafting assignments, and preparing teaching materials such as presentations and quizzes.\\ 
Personalisation    & \student Adapting content to students' individual needs, levels, and preferences. \\ 
Coding             & \student Providing tools for debugging, understanding coding concepts, breaking down problems, and generating code snippets. Offer feedback on code quality and structure.\\
Topic knowledge    & \student  Explain concepts, simulate discussions, provide examples, or suggest further reading materials.\\ 
Feedback           & \student Provide targeted feedback on academic work, including essays, projects, and coding assignments. Identify areas for improvement, suggest revisions, and help students refine their submissions.\\
Summarisation      &\student Condense academic texts, lecture notes, or research articles.\\ 
Assessment         & \student Help with assessment of student assignments.\\
Search             & \student Find relevant academic resources, explore topics, conduct preliminary research, guide literature reviews by summarising key findings and highlighting important sources. \\

\midrule
\multicolumn{2}{c}{\textbf{Issued Warning and Rules}} \\
Human oversight &  \newspaper \student Ensuring that generated outputs are critically reviewed by a human for accuracy, relevance, and appropriateness of the content. Avoiding full reliance on AI for decision-making or content generation.\\
Declaration of use & \newspaper \student Disclosing the use of AI tools. \\
Document AI use & \newspaper Recording all AI use and experiments in an internal register or otherwise document its use in the newsroom. \\
Factual accuracy & \newspaper\student Advising users to cross-check generated outputs against reliable sources to prevent factual inaccuracies or fabricated information.\\
Bias in AI & \newspaper \student Encouraging users to recognise and mitigate biased/discriminatory outputs when interpreting or using generated content.\\
Privacy and sensitive data & \newspaper \student Advising against inputting personal, confidential, or sensitive information into AI tools, particularly those without organisation-approved licenses. \\ 
Copyright & \newspaper \student Cautioning users about the copyright status of both generated content and input data. Users must avoid infringing intellectual property rights and ensure proper attribution where required.\\
AI literacy training &  \newspaper \student Organisational employees need to improve their understanding of AI tools, including their limitations, ethical considerations, and effective usage. \\ 
No prioritisation  & \student Ensuring that AI tools do not unfairly prioritise certain content, viewpoints, or outputs. 
\\ 

Knowledge cut-off  & \student Warning that AI systems may have outdated knowledge due to their training data's temporal limitations. \\
Persuasiveness     & \student Warning that generated content may appear fluent and convincing, even when it is incorrect or nonsensical. \\ 
Source attribution & \student Warning that generated content does not reference its sources, which could lead to plagiarism issues.\\
Digital inequity   & \student Raising awareness about potentially unequal access/utilisation of AI tools among students or institutions. \\ 
Skills assessment     &  \student Warning that AI use might obscure a student’s true abilities.\\ 
Environment        & \student Raising awareness about the environmental impact of using AI tools, which require significant computational resources and contribute to carbon emissions. \\
Teacher load & \student Recognising the potential increase in workload for educators due to the need to monitor, evaluate, and guide AI use in educational settings, as well as address misuse or quality issues.\\  
\bottomrule
\end{tabular}
\caption{Codes used for annotating the policies of news organisations \newspaper and universities \student. }
\label{tab:coding}
\end{table}

\subsection{Selection of Organisational Policies for Analysis}
\label{sec:method:selection}
\textbf{\textit{News Policies.}}
We select 10 news outlets with publicly available policies on generative AI use from across the EU, Switzerland, and the UK, based on \cite{simon2023policies} and \cite{mediumauthor2023guidelines}, which list news organisations' policies. We select from this list the following news outlets by their online availability and under the criteria that they are based in Europe: The Guardian (UK), ANP (Netherlands), Mediahuis (Belgium/Netherlands), Norwegian Tabloid VG (Norway), Le Parisien (France), Financial Times (FT, UK), Süddeutsche Zeitung (SZ, Germany), Der Spiegel (Germany), Ringier (Switzerland) and the BBC (UK). We translate each of the policies into English using Google Translate if they are not available in English. 
We access the latest version of the guidelines, which is documented along with the links to the policies in App., Table \ref{table-app-news-links}. 

\textbf{\textit{University Policies.}} We select 10 universities from across the EU, choosing one top-ranked university from each country\footnote{\url{https://www.topuniversities.com/europe-university-rankings}} which has publicly available guidelines on generative AI use. The selected institutions include: the Technical University of Munich (TUM, Germany), Delft University of Technology (TU Delft, Netherlands), KTH Royal Institute of Technology (KTH, Sweden), Aalto University (Aalto, Finland), Technical University of Denmark (DTU, Denmark), KU Leuven (KUL, Belgium), Swiss Federal Institute of Technology Zurich (ETH, Switzerland), Charles University (CUNI, Czech Republic), University of Vienna (Vie, Austria), University of Lisbon (UdL, Portugal), University of Oslo (UiO, Norway). To reach our target number of ten universities, we checked 27 in total, discarding 17 that did not have public guidelines (six universities from France, three from Spain, two from Sweden, two from Denmark, two from Italy, one from Finland, and one from Switzerland). The guidelines for Vie, UiO, KUL have been automatically translated. We access the latest version of the guidelines, which is documented along with the links to the policies in App., Table \ref{table-app-uni-links}. 
\subsection{Coding of Organisational Policies}
\label{sec:method:coding}
Given the set of policies in the news organisations and universities described above,  we perform iterative coding of these policies. From the identified categories in these policies, we focus on those mentioned by at least two organisations within the same domain. Additionally, we discuss points raised by individual organisations in sections \S\ref{sec:policies:news} and \S\ref{sec:policies:uni}. Where possible, we merge the codes identified across the two domains. 

Table \ref{tab:coding} outlines the resulting codes and their definitions in the respective domain. The codes are grouped into two broad categories: suggested uses of AI within the domain and warnings or rules/requirements related to its use. We use the \use\ symbol to denote a category that the organisation listed as a suggested use. Among the suggested uses, some organisations apply certain restrictions, e.g. limiting the scenarios or issuing a particular prompt to use. We denote such cases we denote with \lock. Furthermore, organisations provide lists of warnings about potential challenges and risks connected with AI use, which we denote as \risk. Finally, organisations also issue rules and requirements for using AI tools, such as declarations of use, which we denote as \rules. 


\section{AI Policies in News Organisations}
\label{sec:policies:news}
\begin{table*}[!t]
\footnotesize
\centering
\begin{tabular}{lcccccccccc}
\toprule
\textbf{Code} & \textbf{FT} & \textbf{ANP} & \textbf{Guardian} & \textbf{Parisien} & \textbf{Spiegel} & \textbf{SZ} & \textbf{BBC} & \textbf{Mediahuis} & \textbf{Ringier} & \textbf{VG} \\
\midrule
\multicolumn{11}{c}{\textbf{Suggested Uses}} \\
Illustrations/Graphics &  \use &   \use &        - &        \use &       \use &  \use &   - &         - &       - &  \use \\
Image generation &  \rules &   \use &        - &        \use &       \use &  - &   - &         - &       - &  \rules \\
Article generation &  \rules &   \use &        - &        \rules &       \use &  \rules &   - &         - &       - &  - \\
Data analysis &  \use &   - &        \use &        - &       - &  \use &   - &         - &       - &  - \\
Language tool &  - &   \use &        \use &        \use &       - &  - &   - &         - &       - &  - \\
Transcription/Translation &  \use &   \use &        - &        - &       - &  \use &   - &         - &       - &  - \\
Ideas (Content) &  - &   \use &        - &        - &       - &  - &   - &         - &       - &  - \\
Ideas (Marketing) &  - &   - &        \use &        - &       - &  - &   - &         - &       - &  - \\
Content Moderation &  - &   - &        - &        - &       - &  \use &   - &         - &       - &  - \\
\midrule
\multicolumn{11}{c}{\textbf{Issued Warning and Rules}} \\
Human oversight &  \rules &   \rules &        \rules &        \rules &       \rules &  \rules &   \rules &         \rules &       \rules &  \rules \\
Declaration of use &  \rules &   - &        \rules &        \rules &       \rules &  \rules &   \rules &         \rules &       \rules &  \rules \\
Factual accuracy & \risk &  \risk &       \risk &        - &      \risk &  - &   \rules &         - &      \risk &  - \\
Bias in AI & \risk &  \risk &        \rules &        - &       - &  - &   - &         \rules &      \risk &  - \\
Privacy and sensitive data &  - &   - &        - &        - &       \rules &  - &   \rules &         \rules &       \rules &  \rules \\
Copyright &  - &   - &        \rules &       \risk &       - &  - &   \rules &         \rules &      \risk &  - \\
AI literacy training &  \rules &   - &        - &        - &       - &  - &   \rules &         \rules &       - &  - \\
Document AI use &  \rules &   - &        - &        - &       \rules &  - &   \rules &         - &       - &  - \\
\bottomrule
\end{tabular}

\caption{Suggested uses, and issued \textit{warnings and rules regarding the use of AI systems} in news organisations. For each element, we denote whether the corresponding guidelines have suggested using AI for its purposes (\use) (with the requirement for human oversight in most cases); have issued warnings regarding the use of AI (\risk), or have issued official rules regarding the use of AI(\rules). See \autoref{tab:coding} for the meaning of the individual codes.}
\label{table-news}
\end{table*}
We describe the policies of 10 news organisations, as summarised in Table~\ref{table-news} towards their suggested uses of AI and issued warnings and rules w.r.t. AI use in the newsroom.
\subsection{Suggested Uses}
All news outlets surveyed encourage the use of AI in their work; however, they propose different degrees of use and applications. Where in Table~\ref{table-news} outlets do not mention any of the listed codes, they still encourage AI usage but do not explicitly list suggested uses in their policy.
Further, all news outlets require human oversight for all or most AI use as well as labelling the output as AI generated.

The use of AI in news organisations has two directions. First, tooling to help automated processes in the day to day work of journalists such as \textbf{data analysis}, \textbf{language tools}, e.g., grammar correction, and \textbf{transcription and translation} of, e.g., interviews. 
Here, there seems to be very little differences in policies -- either, these use cases are mentioned in the policies as allowed or encouraged, or they are not explicitly mentioned, none of the news organisations forbids this type of AI use. In a similar vein, Süddeutsche Zeitung (SZ) explicitly mention the use of AI for \textbf{content moderation}, showing the wide range of possible AI application in news organisations.
The second set of proposed use cases of AI in news organisations is around the content of the published news, i.e., the use of AI for content production. Here policies diverge.
Interestingly, while many news organisations explicitly allow the use of AI for the generation of \textbf{illustrations/graphics}, Financial Times (FT) and VG prohibit the use of AI for \textbf{image generation}, i.e., the creation and publication of photorealistic images whereas ANP, Le Parisien, and Der Spiegel allow use for image generation.
Likewise, policies differ for \textbf{article generation}, i.e., the use of AI to create full or long parts of texts for news articles. While ANP and Der Spiegel allow article creation under human supervision, Financial Times (FT), Le Parisien, and Süddeutsche Zeitung (SZ) explicitly state that their articles are exclusively written by humans.
In the realm of ideation, only ANP encourages the use of AI to support journalists in finding ideas on headlines, articles, and even sources. The Guardian does not mention idea generation for news content, but they do encourage the use of AI for marketing campaign ideas. \looseness=-1
%

%
\subsection{Issued Warning and Rules}

All news organisations mandate human involvement, emphasising \textbf{human oversight} before any generated content is used or published. Outlets like ANP and Der Spiegel require explicit editorial approval for publishing generated content, while BBC and ANP highlight that AI is a supportive tool rather than an independent generator.
Further, all news organisations agree on the \textbf{declaration of use}, setting standards about the transparency of AI use with policies mandating clear labeling of generated content.

Topics of broader society implications, such as \textbf{factual accuracy} and \textbf{bias in AI}, are often mentioned across the news organisations' policies, but few specific rules are proposed.
Generated text may contain factual inaccuracies (see \autoref{fig:sociotech-challenges}). Many of the news organisations warn about this, but only the BBC addresses this issue by requiring that generated content is assessed for ``accuracy and reliability''.
Similarly, inherent bias of AI systems is mentioned across multiple news organisations' policies, with only The Guardian and Mediahuis proposing concrete measures. The Guardian discusses the need to ``guard against the danger of bias'' in both AI models and their training data, while Mediahuis more generally ``watch out for biases in AI systems and work to address them'', emphasising the need to balance the different objectives of journalists, commercial interests, and the audience.

For journalists, sensitive data handling can be a crucial issue. Four of the news organisations have explicit policies on handling \textbf{privacy and sensitive data}. These include the ban of entering into AI systems sensitive data of, as Ringier states in their policy, ``journalistic sources, employees, customers or business partners or other natural persons''. For the BBC, this extends to using AI systems that respect privacy rights.

Given the issues with copyrighted texts used for AI training (see \autoref{fig:sociotech-challenges}), \textbf{copyright} is a central topic for news organisations. Five out of ten analysed news organisations' policies address this topic. 
Rigier only warns about copyright infringement in the context of entering programming code whereas Le Parisien only warns about copyright infringement in the context of publishing AI generated images. 
Unique to The Guardian is the explicit commitment to fair compensation for data creators whose works contribute to AI models, showcasing a progressive view on data usage and ownership rights.

\textbf{AI literacy training} can better equip journalists to deal with AI. Mediahuis, BBC, and Financial Times (FT) support internal training, fostering newsroom awareness and responsible AI deployment. Those trainings include ensuring accountability for AI development and use, train and qualify staff responsible for AI decision, and newsroom awareness (Mediahuis); clear communication on where AI is used, what data is collected and how it works, and affects both staff and audience (BBC); training on the use of AI for story discovery (FT).       

Some news organisations' policies advocate for documenting all instances of AI usage and experimentation. Der Spiegel uses this information to ``exchange information within the company, with other media and with other partners''. The BBC calls for ``clear accountability'' for the use of AI more generally. However, only the Financial Times (FT) explicitly mentions an internal register to track AI use, highlighting an additional layer of accountability not seen in other policies. 

\section{AI Policies in University Education}
\label{sec:policies:uni}

\begin{table*}[!t]
\centering
\footnotesize
\begin{tabular}{lcccccccccc}
\toprule
 \textbf{Code} & \textbf{UiO} & \textbf{Aalto} & \textbf{TUD} & \textbf{KUL} & \textbf{ETH} & \textbf{TUM} & \textbf{Vie} & \textbf{CUNI} & \textbf{DTU} & \textbf{KTH} \\
\midrule
\multicolumn{11}{c}{\textbf{Suggested Uses}} \\
Self-learning      & \use & \use & \use & -    & -    & \use & \use & -    & \use & \lock \\ 
Course design      & \use & \use & -    & -    & \use & \use & -    & \use & \use & -            \\ 
Personalisation    & \use & \use & -    & -    & \use & \use & \use & -    & \use & -            \\
Coding             & \use & \use & \use & \lock & -    & \use & -    & \use & -    & -            \\
Topic knowledge    & \use & \use & \use & \use & -    & -    & \use & -    & -    & \use         \\
Feedback           & \use & \use & \use & -    & \use & -    & \use & -    & -    & -            \\
Language tool      & -    & \use & \use & \use & \use & -    & -    & -    & -    & \lock \\
Ideas           & \use & \use & \use & \use & \use & -    & -    & -    & -    & -            \\
Summarisation      & -    & \use & \use & \use & -    & -    & \use & \use & -    & -            \\
Translation        & \use & \use & -    & -    & -    & -    & \risk& \use & -    & -            \\
Assessment         & \use & -    & \use & -    & \lock & -   & -    & \use & -    & -        \\
Search             & \lock & -   & -    & \use & -    & -    & -    & \use & -    & -            \\
Image generation       & -    & -    & -    & \use & \risk& -    & -    & \use & -    & -            \\

\midrule
\multicolumn{11}{c}{\textbf{Issued Warning and Rules}} \\
Teacher restrictions & \rules & \rules & - & \rules & \rules & \rules & \rules & \rules & - & - \\
AI literacy training & \risk \use      & - & \use & \risk \use & \use & \risk \use & \risk \use & \use & - & -            \\
Human oversight   & \rules     & \rules & - & \rules & \rules & - & \rules & \rules & \rules & \rules \\
Privacy and sensitive data            & \rules      & \risk/\rules & \risk & \risk/\rules & \rules & \risk & \risk & \risk/\rules & \risk/\rules & -            \\
Factual accuracy   & \risk       & \risk    & \risk     & \risk        & \risk  & \risk & \risk & -            & \risk        & \risk        \\
Declaration of use & -           & \rules       & \risk/\rules & \rules & \rules & -     & \rules & \rules & \rules & \rules  \\
Copyright          & \risk/\rules& -            & \risk & \risk/\rules & \rules & \risk     & \risk/\rules & \risk/\rules & \risk & -      \\
Bias in AI      & -           & \risk        & -     & \risk        & \risk  & \risk & -     & -     & \risk & -      \\
No prioritisation  & \risk       & -            & -     & \risk        & -      & \risk & \risk/\rules & -     & -     & \risk  \\
Digital inequity   & \risk       & -            & \risk & \risk        & -      & -     & -     & \risk & -     & -      \\
Knowledge cut-off  & \risk       & \risk        & \risk     & \risk            & -      & \risk & -     & -     & -     & -      \\
Persuasiveness     & \risk       & \risk        & \risk     & -            & -      & \risk & -     & -     & -     & \risk      \\
Source attribution & -           & \risk        & \risk     & -            & \risk  & \risk & \risk     & -     & -     & -      \\
Skills assessment     & \risk       & -            & \risk & -            & \risk  & \risk & -     & -     & -     & -      \\
Environment        & \risk       & -            & \risk & \risk        & -      & -     & -     & -     & -     & -      \\
AI over-reliance       & \risk       & -            & -     & -            & -      & -     & \risk & -     & -     & -      \\
Teacher load & -           & -            & \risk  & -            & -      & -     & -     & -     & \risk     & -     \\  
\bottomrule
\end{tabular}
\caption{Suggested uses, and issued \textit{warnings and rules regarding the use of AI systems} in teaching and learning activities. For each element, we denote whether the corresponding guidelines have suggested using AI for its purposes (\use), have enforced restrictions on the use of AI for it (\lock), have issued warnings regarding the use of AI (\risk), or have issued official rules regarding the use of AI(\rules). See \autoref{tab:coding} for the meaning of individual codes.}
\label{table-uni}
\end{table*}

Table \ref{table-uni} presents a summary of the suggested use cases of AI in teaching and learning as well as the risks and rules enacted in the corresponding institutions. The table presents the points raised by at least two universities, we also discuss points raised by individual universities in the following section.

\subsection{Suggested Uses}

The suggested uses of AI systems in education are framed from two primary perspectives: that of the teacher and the student. The most commonly proposed application is for \textbf{self-learning}, mentioned in seven organisational guidelines. Examples of self-learning activities supported by AI systems include providing students with learning materials and resources, assisting in planning and monitoring their learning, encouraging exploration of covered topics (TUM), and offering alternative perspectives (Aalto).

From the student perspective, AI systems are also frequently viewed as tools for enabling \textbf{personalised learning}, e.g. by recommending individualised learning plans, presenting material with explanations of varying difficulty levels, and enhancing accessibility for students with disabilities (ETH). Additionally, AI is suggested as highly effective for \textbf{coding tasks}, such as understanding concepts, breaking down problems into smaller parts, practising debugging skills, and receiving feedback on code. University guidelines further note that students can leverage AI to \textbf{gain an initial overview of a topic}, terms, or concepts. Moreover, AI can help students monitor their progress and \textbf{provide targeted feedback} on written work or ideas.

AI systems are also commonly suggested as \textbf{language tools}, particularly for grammar, spelling, and reference checks—often considered the safest and most permissible use of AI. For example, KTH identifies this as one of the three approved uses of AI, typically not requiring documentation in the declaration of use. Additional proposed applications for students include \textbf{idea generation, summarising academic literature, translation, student assignment assessment, search engine functionality, and image generation}. However, regarding translation, Vie cautions that AI may produce inaccuracies, particularly with new terminology or less common language combinations.

From the teacher's perspective, suggested uses of AI extend to \textbf{course design} activities, such as formulating learning objectives, drafting rubrics, planning workshops, designing assignments, creating questions, preparing courses, and even simulating a test student. Regarding assessment, ETH explicitly disallows the fully automated grading of student work, but most universities grant teachers the discretion to decide whether \textbf{AI use is restricted} in assignments and exams. \looseness=-1

While many universities provide lists of potential AI applications in teaching and learning, some impose restrictions on those applications. For instance, KTH limits AI usage to predefined prompts and prohibits directly asking the system to generate specific answers or complete assignments. Similarly, KUL restricts the use of AI in coding tasks to generating components of larger assignments, and only if explicitly approved by the teacher.

\subsection{Issued Warnings and Rules}
The most common warnings and rules in university policies regarding AI include considerations of \textbf{privacy} and \textbf{sensitive data}, \textbf{copyright}, \textbf{factual accuracy}, and the \textbf{declaration of use}. While some universities encourage students and teachers to consider privacy and copyright concerns and warn of potential violations, others enforce strict rules regarding the types of data that can be input into AI systems to prevent such issues. For example, Aalto specifies that teachers may only submit student work to university-approved systems and prohibits entering other students' answers or personal information into external systems. To support these policies, Aalto, along with DTU, KUL, and ETH, provides access to Microsoft Copilot for both teachers and students, which is meant to ensure that submitted data is not stored or used for future model training.
Regarding copyright, universities caution that AI can reproduce copyrighted material without proper acknowledgement. They also require users to avoid inputting proprietary information, such as the university’s intellectual property, into AI tools.

Other warnings address risks associated with the quality of generated content, including \textbf{biases} in the content, \textbf{lack of prioritisation} of arguments, \textbf{absence of source attribution} (making verification of accuracy, plagiarism, etc., difficult), \textbf{limited knowledge due to cut-off dates}, and a \textbf{persuasive tone} that can mislead users about the correctness of the information. Risks also arise in educational scenarios where AI integration might \textbf{necessitate course reorganisation or additional assessments} to ensure learning objectives are met and student skills are accurately evaluated. AI usage may also contribute to \textbf{digital inequity}, stemming from disparities in access to paid versus free tools and variations in the quality of generated content based on user skills. \textbf{Over-reliance on AI tools} is another identified risk. \looseness=-1

KUL explicitly highlights the lack of reproducibility as a concern, noting that output can vary between attempts. UiO is unique in warning that AI can produce inappropriate or offensive content. KUL also advises against ``humanising'' AI, emphasising that it is merely a tool.

Many universities stress the \textbf{importance of asking the right questions} and not settling for the first answer provided by AI. To support this, they offer guides for crafting effective prompts, and some institutions even provide dedicated courses on this topic. Teachers are encouraged to \textbf{set clear restrictions on AI use} within their courses and must communicate allowed uses transparently. Finally, most guidelines underscore that both teachers and students remain \textbf{fully responsible} for the content they incorporate into their work, regardless of AI use.

\section{Insights from Organisation Policies for the EU AI Act}\label{sec:insights}
We now put the above findings from organisational policies in the perspective of EU AI Act \cite{2024-ai-act} (\aia). We recognise that these efforts towards AI governance are fundamentally different in scope and process through which they were created, and they are complementary. However, given the ongoing effort\footnote{\url{https://artificialintelligenceact.eu/ai-act-implementation-next-steps/}} to develop a more specific implementation guidelines for \aia, we believe that the insights from the bottom-up policies could help to identify areas where more clarity would be appreciated.

%

\subsection{News Organisations' Policies}
\textit{\textbf{Similarities.}}
News organisations' policies align with the provisions introduced by the \aia for example on the emphasises on \textit{human oversight} with the \aia Article 14.
Article 50 of the \aia describes transparency obligations for providers and deployers, aligning with the requirements to \textit{disclose AI-generated content} by news organisations.
For high-risk AI systems, Article 10 of the \aia regulates \textit{data governance}, a question that also is relevant to downstream users such as journalists. In the context of the \aia this is limited to training, validation, and test data, whereas for news organisations the question of input of sensitive data into AI systems is crucial to preserve privacy of possible sources, reflecting the broader GDPR compliance required under the \aia.
%
%

\textit{\textbf{Gaps.}}
The news' AI policies cover several points not covered by the \aia. In particular, The Guardian emphasises \textit{compensating those whose data is used for AI}, while the \aia lacks explicit provisions for data creators' compensation outside of existing copyright regulations. 
Financial Times and Mediahuis include newsroom \textit{AI training} in their policies. The EU Act currently only requires AI literacy training for developers and deployers of AI systems. If news organisations as downstream users of these systems are not considered as deployers, this requirement will not cover this type of distribution of AI-generated content. 
Multiple policies require addressing \textit{bias in the AI systems} used by journalists, a topic that is yet to be comprehensively covered by legislation.
The \textit{internal AI usage register policy} by Financial Times is an additional documentation practice not specified by the Act but useful for accountability. It could enable retroactive checks on which content was created with which AI system and where the systems where used, which would improve transparency and accountability of this outlet, and hence potentially increase trust in it. 

\subsection{University Education's Policies}
\textit{\textbf{Similarities.}}
The university AI guidelines align with several provisions in the \aia, particularly in their focus on \textit{transparency, human oversight, and data privacy}. Concerns about privacy and data protection, emphasised by universities like KUL and the UiO, align with Article 10 of the \aia, which mandates data governance and safeguards for personal data used for training, testing, and validation in AI systems. Universities advise students and teachers against uploading sensitive data to AI tools, reflecting the broader GDPR compliance required under the \aia.\looseness=-1

The \textit{human oversight of AI} in decision-making, particularly in \textit{assessments and exams}, is echoed in Article 14 of the \aia, which mandates human monitoring of AI decision-making processes in high-risk AI systems. Some universities, such as the TUM, prohibit AI in exams, while others integrate AI tools under strict human oversight, ensuring that AI does not replace independent student evaluation.

The \aia defines education as one of the high-risk areas for AI applications in Article 6 Annex III, enforcing \textit{bias mitigation and fairness obligations}. Universities also raise concerns about AI’s potential to reinforce grading biases and manipulation tactics in AI-driven assessments (e.g., response length bias, goal hijacking). Similarly, universities explore AI-driven adaptive learning and dropout risk prediction (though the latter is clearly a high-risk application that comes with extra considerations for the trade-off between improved performance of Transformer-based systems and the difficulty of regulatory compliance \cite{NielsenRaaschou-PedersenEtAl_2024_Trading_off_performance_and_human_oversight_in_algorithmic_policy_evidence_from_Danish_college_admissions}).

\textit{\textbf{Gaps.}}
University AI policies introduce measures that are either absent or not explicitly covered by the \aia. For example, institutions like UiO, TUD, and KUL highlight the \textit{environmental impact} of AI, focusing on sustainability and ethical AI use. 
Furthermore, university AI policies highlight the risk of increasing \textit{digital inequity} among students, which stems both from disparities in access to paid versus free tools and variations in the quality of generated content based on user skills. The \aia, while regulating AI safety and robustness, does not directly address AI’s carbon footprint, resource consumption, or exacerbated digital inequity.

While the \aia, Article 4 mandates \textit{AI literacy training }for developers and deployers, university policies extend this responsibility to educators and students. Educators are encouraged to equip students with the skills needed to critically evaluate AI outputs. Additionally, both educators and students are provided with resources to learn how to craft effective AI inputs to achieve optimal results. This proactive approach in universities contrasts with the \aia’s narrower focus on professional AI users rather than general AI literacy. 

Overall, while the \aia provides a legal framework for AI safety, transparency, and human oversight, universities take a context-specific approach to AI governance, addressing academic integrity, assessment reliability, and pedagogical challenges in ways that current EU regulation does not yet fully capture.

\subsection{Known research challenges not covered in \aia or organisational policies}
\label{sec:challenges:discuss}

Finally, let us consider the set of sociotechnical challenges that is more broadly known from the existing academic literature (\autoref{fig:sociotech-challenges}), but that we have not found to be addressed in sufficient detail in either organisational policies we considered or \aia:

\textit{\textbf{Definition.}} The organisational policies do not typically define what kind of `AI' is being addressed, and the definition proposed in \aia has been criticized by researchers \cite{Hooker_2024_On_Limitations_of_Compute_Thresholds_as_Governance_Strategy}.

\textit{\textbf{Enforceability.}} Many policies we considered require \textbf{declaration of AI use}, yet there are no robust detection mechanisms to verify compliance \cite{PuccettiRogersEtAl_2024_AI_News_Content_Farms_Are_Easy_to_Make_and_Hard_to_Detect_Case_Study_in_Italian}.

\textit{\textbf{Unsafe outputs.}} Only one university in our sample (UiO) mentioned the possibility of exposing students to inappropriate outputs from AI models.

\textit{\textbf{Misleading marketing claims.}} AI providers are constantly advertising new models claimed to be ever better at `reasoning', `understanding' and other constructs of questionable validity for the current AI \cite{Mitchell_2021_Why_AI_is_Harder_Than_We_Think}. Many policies we examined seem to be influenced by `fear of missing out', manufactured by such narratives. More stringent requirements of transparency for claimed benchmark results could alleviate this problem.

\textit{\textbf{Explainability.}} We saw no policies directly addressing the fact that the current AI systems are not interpretable, which has implications for their use (especially where decisions could have significant consequences, e.g. student grading or news fact-checking).

\textit{\textbf{Brittleness.}} Some university policies mention the need for AI literacy training, but we did not find that in news, and \aia also does not discuss that for the users of AI systems.

\textit{\textbf{Creativity.}} It is possible that over-reliance on AI systems could damage basic competences or creativity of its users, but most policies we examined do not seriously consider this factor.

\textit{\textbf{Carbon emissions.}} Only 3 universities and no news organisations considered this point, and it is not addressed in \aia beside documentation.

\section{Policy Recommendations Based on Gaps Identified in This Work}
To reiterate, while the \aia and organisational policies from universities and news organisations operate within different scopes and serve distinct purposes, they can still inform each other, and insights from academic research can further identify areas not sufficiently addressed by either efforts. This section lists the areas where we believe further governance and research efforts are needed the most. 

%

\textbf{\textit{Expanding AI literacy training.}} The \aia mandates AI literacy training for developers and deployers (Article 4), but it does not mandate AI literacy for students, teachers, journalists, the general public who generate or interact with AI-generated content, or even the media professionals or other users of AI models distributing their outputs on a large scale. Universities integrate AI literacy into academic policies, requiring students and educators to develop critical engagement with AI tools. Institutions like CUNI make a proactive step in this direction emphasising the education in AI ethics and responsible use. Newsrooms such as Financial Times and Mediahuis provide AI training for journalists, ensuring that AI-generated content is fact-checked and responsibly handled. Such training should include also critical reflection on the real functionality of the current AI models vs the marketing hype. While it is important to keep responsibility with the AI developers and deployers, supporting users on how to approach AI will be critical. 

\textit{\textbf{Policies addressing Digital Inequity.}} The \aia does not explicitly address digital inequity, despite its potential to exacerbate social and economic disparities (e.g. due to unequal access to AI and different quality of the models available for different socio-economic and linguistic groups). %
This is particularly relevant in education, where university AI policies have highlighted concerns about disparities in access to paid vs. free AI tools, as well as differences in the quality of AI-generated content based on user skills. Other concerns include the temptation to use AI `education' as a low-cost solution substituting human teachers for the already underpriviledged groups, and siphoning of public resources to for-profit AI providers instead of building public AI infrastructure. All this requires more thought to develop more equitable education infrastructure and policies that consider socio-economic impact on various population groups. 

\textbf{\textit{Improved Transparency for Generated Content.}} While \aia Article 52 mandates disclosure of AI-generated content, it does not specify how AI-assisted work should be attributed or audited by downstream users, or how the record of AI assistance should be kept. Universities enforce strict AI citation rules, requiring students to disclose AI-generated content to uphold academic integrity. However, there is no standardised framework for disclosing AI use, which could aid AI literacy across sectors. For example, in student submissions (e.g. should it be a brief description, or a full prompt+output? How should the source system be specified?) An interesting practice is the internal AI usage registers in Financial Times, which allows editors to track which articles were AI-assisted. 

\textbf{\textit{Getting specific about `bias'.}} Both news and university policies sometimes warn about the possibility of `bias', but it is not clear what kinds of bias should be addressed, or how. This is a gap legislation could address by providing a more comprehensive guidance (e.g. based on existing human rights and non-discrimination laws) for model providers, deployers, as well as downstream users and content distributors.

\textbf{\textit{Attribution and compensation of sources of AI training data.}} EU has copyright laws, but AI training data poses new challenges currently tested in both US \cite{Vynck_2023_Game_of_Thrones_author_and_others_accuse_ChatGPT_maker_of_theft_in_lawsuit,GrynbaumMac_2023_Times_Sues_OpenAI_and_Microsoft_Over_AI_Use_of_Copyrighted_Work} and in Europe\cite{Smith_2024_GEMA_Files_Copyright_Lawsuit_Against_OpenAI_in_Germany}. In our sample, only The Guardian advocates for compensation of content creators whose data is used as part of AI training. 
In education, an equally important factor is source attribution, without which the students could be unwittingly plagiarising existing scholarly work. 
The question of data governance and compensation should be further investigated, taking into account concepts such as data collectives \cite{DBLP:conf/cscw/HsiehZKRDMGEZ24}.

\textbf{\textit{Disclosing Environmental Impact of AI}} The Act does not explicitly address the carbon footprint of AI models besides documentation, despite researchers' concerns about large-scale computational demands \cite{dodge2022measuring,bouza2023estimate,luccioniCountingCarbonSurvey2023,liMakingAILess2023}. Some universities and news outlets highlight the environmental costs of training and running large models, yet there are no regulatory incentives to optimise for sustainability. One ongoing research direction is developing more efficient models \cite{trevisoEfficientMethodsNatural2023}, but if the more efficient models just get used more, this will not help. Mandating a visible disclosure to the users of how much water and energy each AI query costs, and where the resources come from, could help to discourage excessive use. Organisations could also have AI use by their employees as a part of their sustainability reporting. 

\textbf{\textit{Detection and enforceability.}} There are currently no reliable methods of detecting generated text, which makes any policies unenforceable. A promising solution is watermarking \cite{jiang2024watermark,roman2024proactive}, but the providers of commercial LLMs have no incentive to provide a mechanism that could reduce the usage of their services \cite{davisOpenAIWontWatermark2024,gloaguenBlackBoxDetectionLanguage2024}. This is where the considerations of social impact \cite{solaiman2023evaluating} should guide regulation mandating such transparency from the popular AI service providers. 

\textbf{\textit{Clarifying liability.}} In compliance with \aia, providers of AI models may attempt to build in ``safeguards'' to avoid e.g. toxic outputs, and they will have to pass some evaluations to put the model on the market. But should something go wrong, e.g. seriously impacting the mental health of the user, it is currently not clear how much legal recourse the affected users would have. The question of AI liability \cite{Liability_Rules_for_Artificial_Intelligence_European_Commission} will get more pressing with the amount of cases that pose the question of responsibility for the consequences of AI use \cite{awfulai}.

Finally, we would list the following public-interest areas with potential regulatory significance, which need much more research: detection of synthetic content, model interpretability, source attribution to training data, and long-term effects of AI `assistance' on productivity and skills of the workforce. Besides the above suggestions for regulation, these directions should be among priority areas for academic research funding allocation, as the incentives for working on them are just not present in industry.

%




%
%

\section{Conclusion}
In conclusion, the rapid adoption of AI technologies across diverse domains has exposed significant gaps in governance, with multiple organisations scrambling to quickly develop their policies. Our comparative analysis of AI guidelines in universities and news organisations highlights both shared and domain-specific challenges, such as the need for clear accountability mechanisms, addressing biases, and managing domain-specific applications like personalised learning and content generation. We have also identified multiple challenges that are known in academic research, but not addressed by the current policies. These findings underscore the fragmented nature of current governance efforts and the critical need for cohesive policies that balance local organisational needs with broader societal and technological imperatives, while recognizing and supporting areas where more research is needed for better policies. 
By identifying these gaps and challenges, this paper offers actionable insights for refining international legislation and informs the critical future directions of research. Ultimately, bridging the disconnect between local practices, academic research, and global standards is essential for ensuring the safe, fair, and effective deployment of AI across diverse contexts.

\section*{Ethical Considerations Statement}
This study analyses AI policies from ten news organisations and ten universities to identify gaps in the EU AI Act that could be clarified for its implementation and point out possible research directions. All data used in this research is derived from publicly available policy documents, ensuring transparency and compliance with ethical research standards. We strived to present any interpretations or critiques of the policies in a constructive manner to inform policymakers, AI practitioners, and institutional stakeholders.

This study strives to respects intellectual property rights by citing all sources and representing policy content appropriately. Since the analysis pertains to institutional policies rather than individual data, no personally identifiable information is processed or collected.

Finally, we acknowledge that AI governance is an evolving field. To the best of our knowledge, our findings reflect the state of AI policies at the time of analysis and should be interpreted in light of ongoing regulatory and institutional developments. We encourage further interdisciplinary dialogue to refine AI governance frameworks in alignment with ethical, legal, and societal expectations.

\section*{Acknowledgements}
This work was supported by a research grant ([VIL60860]) from Villum Fonden (recipient: Anna Rogers). Pepa Atanasova’s work on this paper was supported by The Villum Synergy Programme ([VIL40543]).

\bibliographystyle{ACM-Reference-Format}
\bibliography{main}

\appendix

\section*{Appendix}

\begin{table*}[ht]
\begin{tabular}{lp{10cm}l}
\toprule
\textbf{News Organisation} & \textbf{Policy Link} & \textbf{Version} \\
\midrule
Guardian & \url{https://www.theguardian.com/help/insideguardian/2023/jun/16/the-guardians-approach-to-generative-ai} & June 2023 \\
ANP & \url{https://drive.google.com/file/d/1-3sAJtkOJrdIGw-gZFqYDEGQQNw13e0U/view} & April 2023 \\
Mediahuis & \url{https://www.independent.ie/editorial/editorial/aiframeworkguide140623.pdf} & May 2023 \\
VG & \url{https://www.vg.no/informasjon/redaksjonelle-avgjorelser/188}& April 2023 \\
Parisien & \url{https://www.cbnews.fr/medias/image-groupe-echos-parisien-s-engage-face-intelligence-artificielle-generative-76799} & May 2023 \\
FT & \url{https://www.ft.com/content/18337836-7c5f-42bd-a57a-24cdbd06ec51}& May 2023 \\
SZ & \url{www.ethz.ch/en/the-eth-zurich/education/ai-in-education.html} & Dec. 2024 \\
Spiegel & \url{https://www.sueddeutsche.de/projekte/artikel/kolumne/kuenstliche-intelligenz-ki-e903507/} & June 2023 \\
BBC & \url{https://www.bbc.co.uk/supplying/working-with-us/ai-principles/} & Feb. 2024 \\
Ringier & \url{https://www.ringier.com/ringier-introduces-clear-guidelines-for-the-use-of-artificial-intelligence/} & May 2023 \\
\bottomrule
\end{tabular}
\caption{Links to AI policies and their versions for each news organisation.}
\label{table-app-news-links}
\end{table*}

\begin{table*}[ht]
\begin{tabular}{lp{10cm}l}
\toprule
\textbf{University} & \textbf{Guidelines Link} & \textbf{Version} \\
\midrule
TUM & \url{www.prolehre.tum.de/fileadmin/w00btq/www/Angebote_Broschueren_Handreichungen/ProLehre_Handreichung_ChatGPT_EN.pdf} & Feb. 2023 \\
TU Delft & \url{www.tudelft.nl/teaching-support/educational-advice/assess/guidelines/ai-chatbots-in-unsupervised-assessment} & June 2023 \\
KTH & \url{www.kth.se/profile/wouter/page/chatgpt-pragmatic-guidelines-for-students-september-2023} & Sep. 2023 \\
Aalto & \url{www.aalto.fi/en/services/guidance-for-the-use-of-artificial-intelligence-in-teaching-and-learning-at-aalto-university}& Aug. 2023 \\
DTU & \url{www.dtu.dk/english/newsarchive/2024/01/dtu-opens-up-for-the-use-of-artificial-intelligence-in-teaching} & Jan. 2024 \\
KUL & \url{www.kuleuven.be/english/genai}& - \\
ETH & \url{www.ethz.ch/en/the-eth-zurich/education/ai-in-education.html} & Dec. 2024 \\
CUNI & \url{www.ai.cuni.cz/AI-12-version1-ai_elearning_en.pdf} & June 2023 \\
Vie & \url{www.studieren.univie.ac.at/en/studying-exams/ai-in-studies-and-teaching/} & Sep. 2024 \\
UdL & \url{www.conselhopedagogico.tecnico.ulisboa.pt/files/sites/32/ferramentas-de-ai-no-ensino-v8-1.pdf} & Nov. 2023 \\
UiO & \url{www.uio.no/english/services/ai/} & - \\
\bottomrule
\end{tabular}
\caption{Links to AI guidelines and their versions for each university.}
\label{table-app-uni-links}
\end{table*}

\end{document}